\documentclass[sigconf]{acmart}

\AtBeginDocument{%
  }

\copyrightyear{2025}
\acmYear{2025}
\setcopyright{acmlicensed}\acmConference[UniSearch]{Proceedings of the xxx}{Beijing}{China}
\acmBooktitle{the Preprint Version of UniSearch, Beijing, China}
\acmDOI{10.1145/xxxxx}
\acmISBN{xxxxx/2025/09}

\usepackage{amsfonts,amssymb}
\usepackage{color}
\definecolor{purplefish}{RGB}{138,43,226}
\definecolor{orangefish}{RGB}{210,105,30}
\definecolor{crimson}{RGB}{220,20,60}
\usepackage{mathrsfs}
\usepackage{url}
\usepackage{subfigure}
\usepackage{graphicx}
\usepackage{multirow}
\usepackage{makecell}

\usepackage{algorithmic}
\usepackage{algorithm}
\usepackage{amsmath} 
\usepackage{amsfonts}

\begin{document}

\title{UniSearch: Rethinking Search System with a Unified Generative Architecture}


\author{Jiahui Chen$^{\dagger}$, Xiaoze Jiang$^{*\dagger}$, Zhibo Wang$^{\dagger}$, Quanzhi Zhu$^{\dagger}$,
        Junyao Zhao$^{\dagger}$, Feng Hu$^{\dagger}$, Kang Pan, 
        \\  Ao Xie, Maohua Pei, Zhiheng Qin, Hongjing Zhang, Zhixin Zhai, Xiaobo Guo, Runbin Zhou,  
        \\ Kefeng Wang, Mingyang Geng, Cheng Chen, 
        Jingshan Lv,
        Yupeng Huang, Xiao Liang, Han Li}
\affiliation{%
  \institution{Kuaishou Technology, Beijing, China}
  \country{}
}
\email{{chenjiahui11, jiangxiaoze, wangzhibo07, 
        zhuquanzhi03, zhaojunyao, hufeng}@kuaishou.com}

\begin{abstract}
Modern search systems play a crucial role in facilitating information acquisition. Traditional search engines typically rely on a cascaded architecture, where results are retrieved through recall, pre-ranking, and ranking stages. The complexity of designing and maintaining multiple modules makes it difficult to achieve holistic performance gains. Recent advances in generative recommendation have motivated the exploration of unified generative search as an alternative. However, existing approaches are not genuinely end-to-end: they typically train an item encoder to tokenize candidates first and then optimize a generator separately, leading to objective inconsistency and limited generalization. To address these limitations, we propose UniSearch, a unified generative search framework for Kuaishou Search. UniSearch replaces the cascaded pipeline with an end-to-end architecture that integrates a Search Generator and a Video Encoder. The Generator produces semantic identifiers of relevant items given a user query, while the Video Encoder learns latent item embeddings and provides their tokenized representations. A unified training framework jointly optimizes both components, enabling mutual enhancement and improving representation quality and generation accuracy. Furthermore, we introduce Search Preference Optimization (SPO), which leverages a reward model and real user feedback to better align generation with user preferences. Extensive experiments on industrial-scale datasets, together with online A/B testing in both short-video and live search scenarios, demonstrate the strong effectiveness and deployment potential of UniSearch. Notably, its deployment in live search yields the largest single-experiment improvement in recent years of our product’s history, highlighting its practical value for real-world applications.
\end{abstract}




\maketitle

\section{Introduction}

\begin{figure}[t]
\centering
\includegraphics[width=8.5cm]{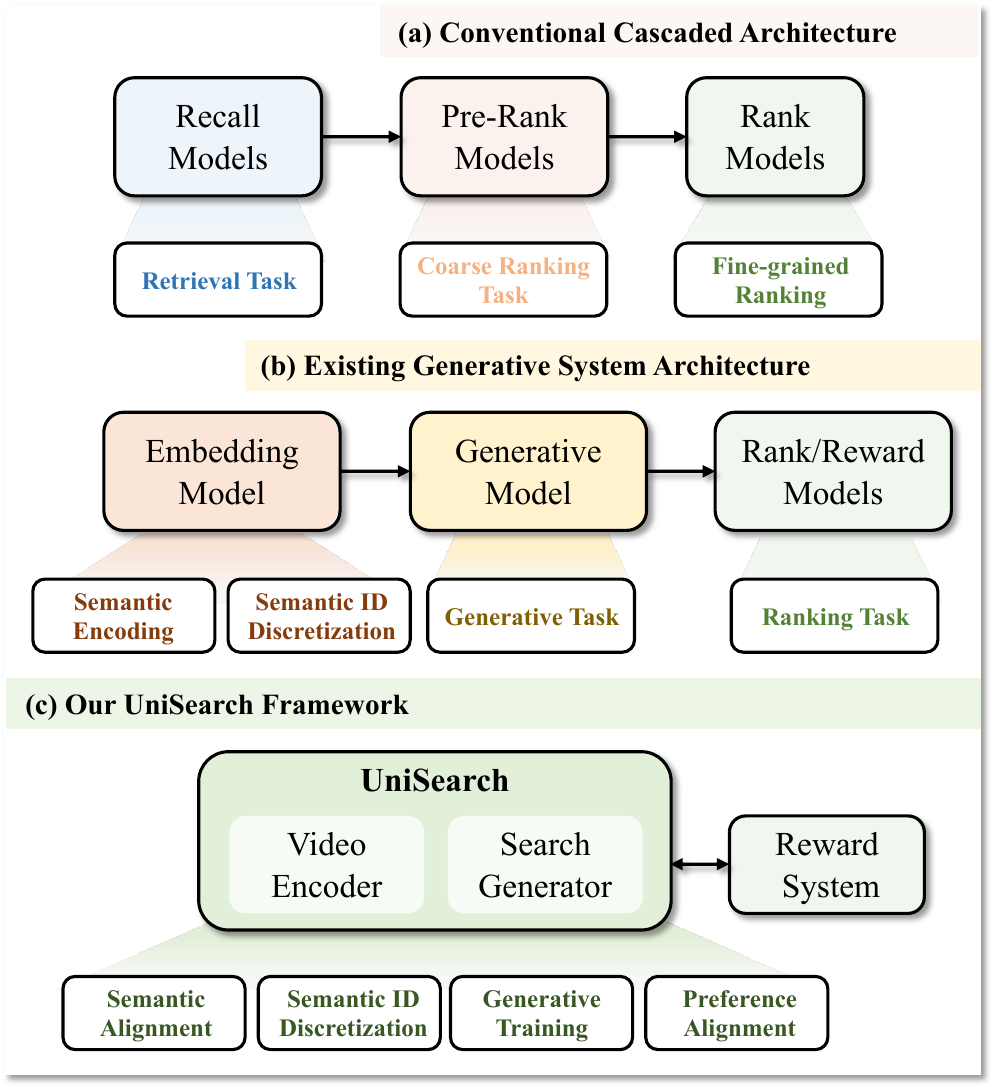}
\caption{A schematic illustration of existing search architecture and our UniSearch. (a) Conventional Cascaded Architecture, (b) Existing Generative System Architecture and (c) Our UniSearch Framework.
}
\label{pic:basicIdea}
\end{figure}

Search systems are fundamental to modern industrial applications and information networks, enabling users to efficiently access relevant content from massive item collections~\cite{bai2025unconstrained,lin2024enhancing}. Over the past two decades, search technology has advanced significantly, evolving from classical inverted indexing to sophisticated deep neural models. Despite these advances, most industrial search engines still adopt a multi-stage cascaded architecture (MCA), which filters candidate items in a coarse-to-fine manner through recall, pre-ranking, and ranking stages~\cite{khandagale2025interactrank}, as shown in Figure~\ref{pic:basicIdea} (a).  

Although this cascaded design has been refined through years of practice, it suffers from several inherent drawbacks. First, each stage employs distinct models with separate optimization objectives, resulting in misaligned training signals across stages~\cite{zheng2024full}. This discrepancy often prevents optimal end-to-end relevance modeling, limiting the system’s upper-bound performance. Second, the multi-stage structure introduces substantial inference latency and high maintenance overhead, as each stage requires dedicated algorithmic design and system support~\cite{deng2025onerec}. These limitations motivate the search for \emph{unified generative solutions} that can replace the cascaded pipeline with a single end-to-end framework.  

The rapid progress of large-scale generative models in recent years has made this direction increasingly viable~\cite{floridi2020gpt,yang2025qwen3,wang2025lad,touvron2023llama,liu2024deepseek}. In recommendation systems, several works have demonstrated that generative models can replace the entire cascaded pipeline with a unified solution, achieving remarkable results~\cite{zhou2025onerec,deng2025onerec}. However, most existing approaches are not truly end-to-end, as shown in Figure~\ref{pic:basicIdea} (b): they typically involve two separate training steps---first training a video encoder to produce embeddings and semantic tokens, followed by training a generator to predict these tokens. This discrepancy between tokenization and generation objectives leads to inconsistent optimization and suboptimal generation quality. More importantly, extending generative models to search poses unique challenges: unlike recommendation, which remains an intra-domain task (mapping user/item embeddings to items), search requires \emph{cross-domain generation} from text queries directly to items. Existing studies have only explored generative models in early stages such as recall~\cite{yang2023auto,liu2024asi++}, and the exploration of unified generative methods as a replacement for the full cascaded search architecture remains underexplored.

To address these challenges, we propose UniSearch, an end-to-end generative search framework deployed at industrial scale. As illustrated in Figure~\ref{pic:basicIdea} (c), UniSearch adopts a unified architecture comprising a \emph{Search Generator} and a \emph{Video Encoder}. The Generator follows an encoder–decoder paradigm, taking query text and user features as input and autoregressively generating the semantic identifiers (SID) of relevant items. The Video Encoder, in turn, learns latent embeddings from multi-modal signals (e.g., textual, visual, and statistical data), and discretizes them into semantic IDs using VQ-VAE~\cite{van2017neural}, thereby constructing a compact representation space that bridges queries and items within a shared generative framework. 
Built upon this architecture, we introduce a unified pre-training strategy that seamlessly integrates semantic tokenization with autoregressive generation. In contrast to prior work that separates item encoding and query generation into independent objectives~\cite{deng2025onerec,guo2025onesug}, UniSearch jointly optimizes both tasks. Specifically, residual contrastive learning, guided by a coarse-to-fine optimization strategy, aligns the semantics between queries and videos. Simultaneously, the discretization process in VQ-VAE produces semantic IDs that are co-optimized with the generative model. This design eliminates the objective mismatch common in previous generative recommendation approaches, enabling mutual reinforcement between video encoding and query generation, and ultimately leading to more robust representations and higher-quality outputs. 
Beyond supervised pre-training, UniSearch incorporates an online post-training phase for \emph{Search Preference Optimization} (SPO). In a controlled online environment, UniSearch generates candidate results that are evaluated through a reward system combining system-estimated signals (e.g., relevance and content quality) with real user feedback, such as clicks and watch time. Leveraging the SPO algorithm, UniSearch refines its generation policy based on these preference signals, aligning its outputs more closely with user intent. 
To ensure efficiency and scalability in real-world deployment, we employ a Trie-based inference mechanism that constrains generation to valid semantic ID paths. This not only prevents invalid outputs but also enhances computational efficiency, making UniSearch practical for large-scale applications. In Kuaishou’s production environment, UniSearch is integrated with a reinforcement learning (RL) training system, an online inference service accelerated by TensorRT and KV-cache, a dynamic Trie server, and a reward feedback system—together forming a complete, end-to-end generative search solution.

We extensively evaluate UniSearch on large-scale industrial datasets and real-world online traffic. Offline experiments demonstrate that UniSearch outperforms strong cascaded baselines in both relevance and efficiency. Online A/B testing on Kuaishou’s live and short-video search systems further confirms its effectiveness.
The contributions of this paper are summarized as follows:

(1) UniSearch is the \textbf{generative search solution deployed at scale}, replacing the traditional cascaded architecture with a unified model that surpasses it in both retrieval quality and computational efficiency.

(2) We propose a \textbf{unified training framework} that jointly integrates tokenization and generation tasks, leading to more effective item encoding and higher-quality generation compared with conventional multi-stage training.

(3) We successfully deploy UniSearch in Kuaishou’s live and short-video search systems. Large-scale online A/B experiments verify its practical effectiveness and efficiency, serving hundreds of millions of active users.

\section{Related Works}

\subsection{Search System}

Traditional search systems are typically built upon a cascaded pipeline architecture, including recall, pre-ranking and ranking. Within this framework, extensive research has been devoted to retrieving query-relevant content across different stages of the pipeline \cite{zhang2022multi,xu2024optimizing,khandagale2025interactrank,xiong2024search,ye2024exploring}. Modern search systems adopt BERT-based \cite{devlin-etal-2019-bert} discriminative models across different stages to better capture relevance, with distinct architectures and training objectives tailored to each phase. In the recall stage, a dual-tower architecture combined with contrastive learning is commonly employed to enhance the model’s semantic retrieval capability. For pre-ranking, a dual-tower model with post-interaction is used to efficiently filter the large set of candidates returned by the recall stage. Finally, the fine-ranking module adopts a single-tower, fully interactive architecture, where point-wise learning is applied to accurately estimate query-document relevance, with a strong emphasis on model precision. This requires sustained and significant algorithmic design and investment. However, the objectives of different components often diverge, potentially leading to suboptimal overall performance. In contrast to prior methods, we use a single generative model to replace the traditional multi-stage cascade, unifying objectives and improving end-to-end consistency.

\subsection{Generative Retrieval and OneModels}
Recently, autoregressive generation-based Large Language Models (LLMs) \cite{openai2025gpt4} have exhibited remarkable capabilities  across multiple domains, such as recommendation \cite{deng2025onerec,zhou2025onerec,zhai2024hstu,han2025mtgr,yang2025sparse,huang2025genrank,zheng2025ega}. This significant breakthrough has led to the emergence of a new paradigm in retrieval \cite{li2025matching,li2024unigen,liu2024asi++,mustar2020using,pang2025GRAM,rajput2023recommender,tang2024list,tay2022transformer,wang2025lad,yin2020learning,zheng2024adapting,li2024GenR-PO} - the shift from matching-based to generation-based approaches, known as generative retrieval. Generative retrieval represents a recent paradigm shift in information retrieval (IR), where retrieval is formulated as a text generation task \cite{li2025matching,li2024unigen}. Instead of retrieving documents by scoring candidates, a generative retriever directly generates document identifiers, passages, or even answers conditioned on the input query \cite{wang2025lad,tang2024list}. In the field of text-to-text information retrieval, such as query auto-completion, recent approaches directly generate the corresponding answer conditioned on the input text \cite{wang2025lad,yin2020learning,mustar2020using}. Compared to dual-encoder (bi-encoder) models, this generative paradigm breaks the limitations of traditional indexing and has achieved significant improvements in performance metrics. Other works in the recommendation domain typically encode documents into discrete paths, which are then used to generate candidates as a novel source for retrieval \cite{tay2022transformer,rajput2023recommender,liu2024asi++,zheng2024adapting,pang2025GRAM}. The above-mentioned studies focus on applying generative models within the retrieval component, while neglecting their potential to substitute the entire cascaded architecture through their strong generative capabilities. By contrast, recent efforts have started to use one generative model as a replacement for the traditional pipeline, such as OneRec on short-video recommendation \cite{deng2025onerec,zhou2025onerec}, OneSug on  query auto-completion \cite{guo2025onesug}, POI recommendation \cite{wang2025onepoi} and local life service \cite{cai2025oneloc}. However, many of these methods rely on pre-tokenized candidate items, which can limit generalization and hinder the model’s ability to adapt to unseen or dynamic content. In contrast, our work introduces a unified generative framework that removes the reliance on fixed candidate tokenization, enabling open-ended generation of item identifiers.

\section{Methodology}

In this section, we first present the overall architecture of UniSearch, followed by its training scheme, which consists of unified pre-training and search preference alignment post-training. Finally, we describe its inference process and deployment.

\subsection{UniSearch Architecture}

\begin{figure*}[t]
\centering
\includegraphics[width=1.0\textwidth, trim=100 40 100 20, clip]{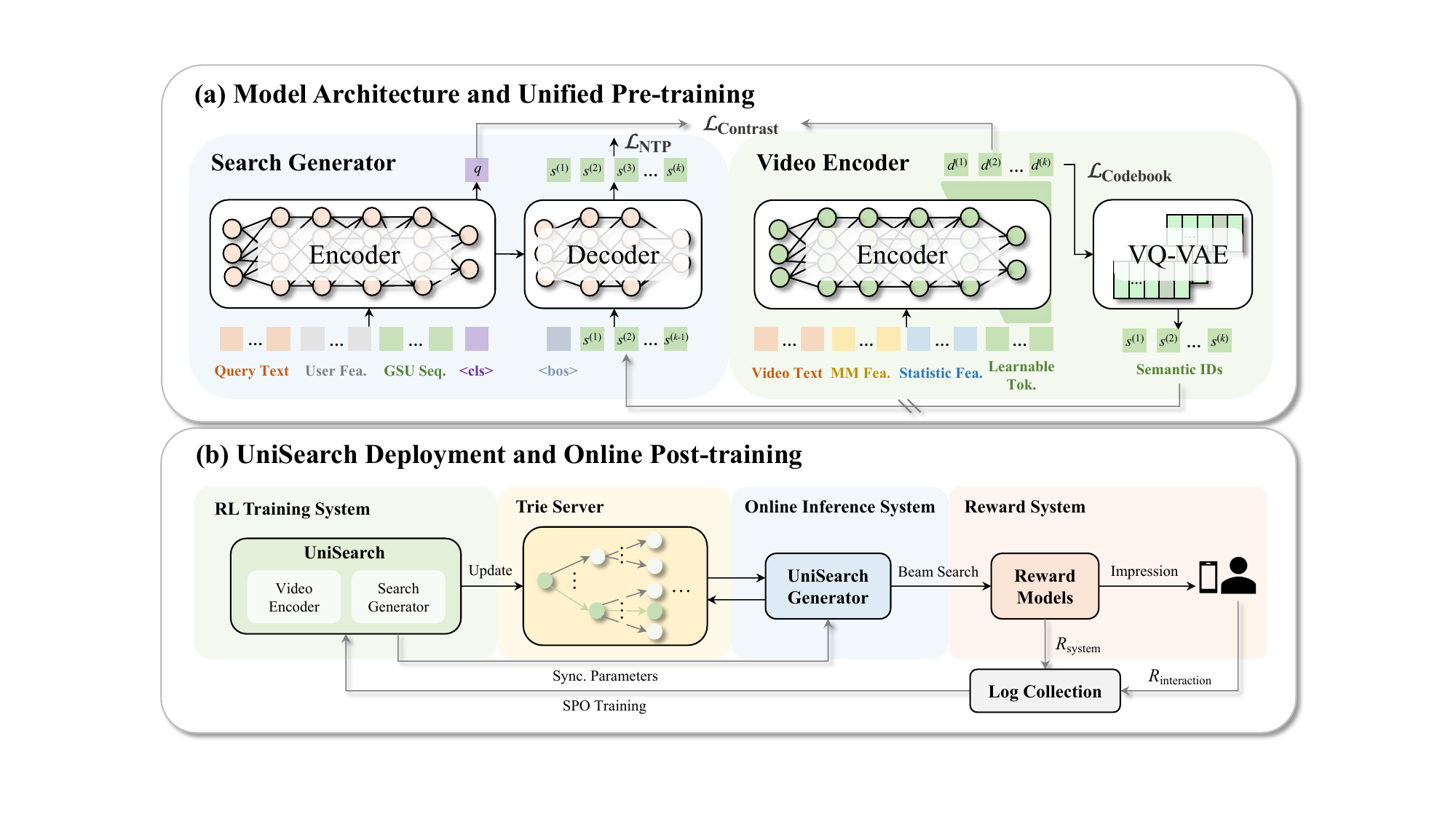}
\caption{The overall architecture of the UniSearch, where ``Seq.'' denotes  ``Sequence'', ``Fea.'' stands for ``Feature'', ``Learned Tok.'' refers to the learned tokens for documents,  and ``MM Fea.'' represents multi-modal features. 
}
\label{pic:model}
\end{figure*}

Unlike previous cascaded search systems that rely on multiple models across different stages, UniSearch adopts a unified architecture to perform both training and inference within a single framework. This design eliminates the inconsistencies between stage-specific objectives and reduces system complexity, enabling more robust and efficient end-to-end learning.  
As illustrated in Figure~\ref{pic:model} (a), UniSearch consists of two main components: a \textbf{Search Generator} and a \textbf{Video Encoder}.  

\noindent\textbf{Search Generator.} 
The Generator follows an encoder--decoder generative paradigm. The encoder is a bidirectional Transformer that processes the user query along with auxiliary user features (\textit{e.g.}, GSU sequences representing personalized historical behaviors). To further capture the holistic semantics of a search request, we prepend a special \texttt{<cls>} token to the input sequence. The resulting representation of \texttt{<cls>} serves as the global query embedding $q$, which summarizes the intent of the current search request. The decoder then receives the contextualized query representation from the encoder, and autoregressively generates a sequence of tokens corresponding to the semantic IDs of the relevant videos. This operation effectively transforms a natural language query into a structured semantic representation of search results.  

\noindent\textbf{Video Encoder.} 
The Video Encoder is a unidirectional Transformer designed to learn latent embeddings and semantic IDs for each video item. Specifically, each video is represented through its textual metadata, multi-modal content features, and side statistical features. These inputs are concatenated with $k$ learnable tokens and passed through the Transformer, producing a sequence of $k$ latent embeddings: $D = \{d^{(1)}, d^{(2)}, \ldots, d^{(k)}\}$, that capture the semantic properties of the video. To discretize these continuous embeddings into semantic identifiers, the Video Encoder is augmented with a VQ-VAE module, which maps the latent embeddings into a codebook space. The resulting semantic ID (SID) sequence is: $S = \{s^{(1)}, s^{(2)}, \ldots, s^{(k)}\}$, which provides a compact and interpretable representation of each video item, ensuring that items can be indexed and retrieved consistently in the generative search framework.  

By jointly employing the Generator and Video Encoder, UniSearch directly bridges the gap between textual queries and item representations, allowing the entire search process to be handled within a single, coherent framework.  

\subsection{Unified Pre-training}

To enable the Generator and Video Encoder to work cooperatively, we design a \textbf{unified pre-training framework} that simultaneously learns video encoding and query-to-video generation. Unlike prior generative recommendation methods that separate item tokenization and generation into two disjoint stages, UniSearch optimizes both objectives in a single training pipeline, ensuring semantic consistency and eliminating task misalignment.  

Our pre-training data are sampled from large-scale search logs, where each record contains:
query, user features, candidate video set $\{d_1, d_2, \ldots, d_N\}$, label set $\{l_1, l_2, \ldots, l_N\}$.
Here, the candidate video set includes both exposed items and unexposed negative items for the query, while the label set provides graded relevance annotations for each candidate, reflecting both query--video relevance and video quality.  

\noindent \textbf{Residual Contrastive Learning for Semantic Alignment.}  
The first objective of UniSearch pre-training is to align the semantics of user queries with the latent representations of relevant items. Let $q$ denote the query embedding produced by the Generator encoder and $D_i$ the latent representation of the $i$-th candidate video encoded by the Video Encoder. To achieve this, we introduce \emph{residual contrastive learning}, where each video embedding is represented as a sequence of residual components $d_{i}^{(n)}$. At step $n$, the current residual $d_{i}^{(n)}$ is optimized against the query embedding $q_i$ while accumulating information from all previous residuals, as defined in Equation~\ref{eq:loss_contrast}. This residual formulation encourages each token to capture complementary information, thereby reducing redundancy across tokens and making more efficient use of the codebook, while also alleviating the token-collapsing issue common in discrete representation learning.  
\begin{equation}
\label{eq:loss_contrast}
    \mathcal{L}_{\text{contrast}} = \sum_{n=1}^{k} \mathcal{L}\Big(q_{i}, \text{sg}\big[\sum_{m<n} d_{i}^{(m)}\big] + d_{i}^{(n)}\Big),
\end{equation}
\begin{equation}
    \mathcal{L}(q, d) = -\log \frac{\exp \big(\text{sim}(q, d) / \tau\big)}{\exp\big(\text{sim}(q, d) / \tau\big) + \sum\limits_{d^{-} \in \mathcal{N}}\exp\big(\text{sim}(q, d^{-})/\tau\big)},
\end{equation}
where $\text{sg}(\cdot)$ denotes the stop-gradient operation, $\tau$ is the temperature parameter, and $\mathcal{N}$ is the negative sample set. In contrast to conventional cosine similarity, we adopt L2 similarity $\text{sim}(q, d) = 1 - \|q - d\|_{2}^{2}$, which better fits the residual accumulation mechanism and preserves the geometric structure of embeddings.  

\noindent \textbf{Coarse-to-Fine Strategy.}  
To further strengthen semantic alignment, we design a coarse-to-fine training strategy that mimics the cascaded ranking process in traditional search systems. The intuition is that the first residual token $d^{(1)}$ should learn an easy, coarse-grained matching task, ensuring robust recall capability by distinguishing relevant from irrelevant items at a broad level. Subsequent tokens $d^{(2)}, \ldots, d^{(k)}$ progressively take on more complex tasks, refining the representation toward fine-grained ranking and personalization. This staged learning process is achieved by gradually increasing the difficulty of the negative sample set $\mathcal{N}$: $d^{(1)}$ is trained against simple in-batch negatives, while later residuals are optimized against increasingly hard negatives sampled from semantically similar but irrelevant items.  

Compared with residual quantization methods such as RQ-Kmeans that perform two-stage residual clustering, our approach enables end-to-end training, thereby reducing clustering loss and avoiding the mismatch between separate training objectives. The residual formulation not only decomposes the learning task into a sequence of increasingly difficult subtasks but also prevents token redundancy, making more effective use of the codebook. As a result, UniSearch simultaneously captures strong recall ability and refined ranking capacity within a single unified framework, achieving better overall search performance than approaches that optimize only for recall.

\noindent \textbf{Discretization into Semantic IDs.} 
While semantic alignment ensures meaningful item representations, the Video Encoder must also discretize latent embeddings into semantic IDs for generative modeling. We employ VQ-VAE \cite{van2017neural} for this purpose. Unlike prior works that rely on post-clustering methods such as balanced K-means or RQ-Kmeans~\cite{deng2025onerec, guo2025onesug}, VQ-VAE updates the codebook jointly during training, enabling efficient online learning and inference. Concretely, for each embedding $d_{i}^{(n)}$, the VQ-VAE encoder performs nearest-neighbor lookup in the learnable codebook to obtain a quantized embedding $e_{i}^{(n)}$ and corresponding semantic ID $s_{i}^{(n)}$. The quantized embeddings $E_{i} = \{e_{i}^{(1)}, e_{i}^{(2)}, \ldots, e_{i}^{(k)}\}$ are further required to reconstruct the original embeddings $D_{i}$. 
Codebook loss is introduced to ensure $e_{i}^{(n)}$ approximates $d_i$, as shown in Equation~\ref{eq:loss_codebook}:
\begin{equation} \label{eq:loss_codebook}
    \mathcal{L}_{\text{codebook}} = \sum_{n=1}^{k} \alpha_{1}||\text{sg}[d_{i}^{(n)}] - e_{i}^{(n)}||_{2}^{2} + \alpha_{2}||d_{i}^{(n)} - \text{sg}[e_{i}^{(n)}]||_{2}^{2}.
\end{equation}
Specifically, we employ the SimVQ strategy~\cite{zhu2024addressing} to further stabilize the discretization process and avoid codebook collapse. With the VQ module, videos can be discretized into semantic IDs in a fully end-to-end manner, without the inconsistencies of offline clustering.  

\noindent \textbf{Reject-Sampling Generative Training.} 
Once items are mapped into semantic IDs, the Generator is trained to produce these sequences autoregressively using a next-token prediction objective with cross-entropy loss. To enhance generation quality, we employ reject-sampling strategies during training: low-quality items (as determined by labels) are filtered out, and losses for items of different levels are reweighted accordingly, as shown in Equation~\ref{eq:loss_ntp}:
\begin{equation} \label{eq:loss_ntp}
    \mathcal{L}_{\text{NTP}} = -w_{i}\sum_{n=1}^{k}\log p\big(s_{i}^{(n)}|q, u, s_{i}^{(<n)} \big),
\end{equation}
where $w_{i}$ is a re-weight factor determined by label $l_{i}$.
This ensures that the Generator prioritizes high-relevance, high-quality items when producing search results.  

\noindent \textbf{Overall.} 
Through the above unified pre-training objectives, Uni-Search achieves joint optimization of query--item semantic alignment, item discretization, and high-quality sequence generation. This provides a consistent and robust foundation for subsequent post-training and inference. The overall training loss is as follows:
\begin{equation} \label{eq:loss_total}
    \mathcal{L} = \lambda_{1} \mathcal{L}_{\text{contrast}} + \lambda_{2} \mathcal{L}_{\text{codebook}} + \lambda_{3} \mathcal{L}_{\text{NTP}},
\end{equation}
where $\lambda_{1, 2, 3}$ are hyper-parameters to balance the loss scale and stabilize training process.

\subsection{Post-training with SPO}

After pre-training, UniSearch is already capable of generating usable search results. However, to further enhance the quality of generated results and better align them with user preferences, we introduce an online reinforcement learning framework termed Search Preference Alignment. This post-training stage is conducted in a controlled online environment, as illustrated in Figure~\ref{pic:model} (b).  

Specifically, the deployed UniSearch model (see Section~\ref{sec:infer} for details) generates $N$ candidate result sequences using Beam Search. These candidates are passed to a Reward System, which assigns each result a reward score based on multiple factors such as relevance and video quality. The top-$M$ results are then exposed to users. Subsequently, users' real interaction behaviors (\textit{e.g.}, clicks, likes, or downloads) are also collected and incorporated as an additional reward signal. The combined reward scores are used to update UniSearch through the Search Preference Optimization (SPO) algorithm, thereby aligning the model’s output distribution with user preferences in an online, data-driven manner.  

\noindent \textbf{Reward System.}  
Modern industrial search engines have undergone years of refinement, particularly in ranking modeling. To leverage this, we adopt the existing fine-ranking module of the production search system as the Reward System. This module takes as input the query, user context, and candidate video features, and outputs multiple predictive scores covering dimensions such as query--item relevance, video quality, and expected user satisfaction. These scores constitute part of the reward signal. In addition, once the videos are presented to the user, explicit interaction feedback (\textit{e.g.}, click-through, watch time, likes, or downloads) provides another intuitive measure of search quality. The final reward score is a weighted combination of system-estimated scores and observed user interactions:
\begin{equation} \label{eq:reward}
    R = \gamma_{1} R_{\text{system}} + \gamma_{2} R_{\text{interaction}},
\end{equation}
where $\gamma_{1, 2}$ balance the relative importance of system-predicted and user-behavioral signals.

\noindent \textbf{Search Preference Optimization.}
From the online system logs, we obtain $G$ generated search results under a given query, denoted as $\{S_1, S_2, \ldots, S_G\}$, together with their corresponding reward scores $\{R_1, R_2, \ldots, R_G\}$. For each candidate $S_i=\{s_{i}^{(1)}, s_{i}^{(2)},...,s_{i}^{(k)}\}$, UniSearch assigns a generation probability $p_i = \pi_\theta(S_i \mid q, u)$ conditioned on the query $q$ and user context $u$.  
Following GRPO \cite{liu2024deepseek}, we first compute the relative advantage of each candidate by normalizing its reward against the expected reward of the batch:  
\begin{equation} \label{eq:advantage}
A_i = \frac{R_i - \text{mean}(\{R_1, R_2, \ldots, R_G\})}{\text{std}(\{R_1, R_2, \ldots, R_G\})}.
\end{equation}
This formulation encourages the model to prioritize results that are better than the batch average while suppressing worse ones.  
The final optimization objective is then defined as:  
\begin{align}
L_{SPO}(\theta) = &-
{\frac{1}{G}}{\sum_{i=1}^{G}}{\frac{1}{k}}{\sum_{n=1}^{k}}\Big({\frac{\pi_\theta\big(s_{i}^{(n)}|q,u,s_{i}^{(<n)}\big)}{\pi_{\theta_{\text{no\_grad}}}\big(s_{i}^{(n)}|q,u,s_{i}^{(<n)}\big)}}A_i
\nonumber
\\
&-\beta\mathbb{D}_{\text{KL}}[\pi_\theta||\pi_{ref}]\Big),
\end{align}
where the first term maximizes the likelihood of preference-aligned generations weighted by their relative advantages, and the second term regularizes the updated policy $\pi_\theta$ against the reference policy $\pi_{\text{ref}}$ obtained from pre-training. The coefficient $\beta$ controls the strength of the KL regularization.  

This formulation enables UniSearch to align its generation behavior with both system-estimated and user-observed preferences, while maintaining stability and avoiding catastrophic deviation from the pre-trained model.

\subsection{Inference and Deployment} \label{sec:infer}

To improve efficiency during inference, we constrain the generation of UniSearch with a prefix tree (Trie) structure that encodes all valid semantic ID sequences. A dedicated Trie service maintains the dictionary of permissible paths, ensuring that UniSearch only generates valid sequences. At each decoding step, UniSearch queries the Trie service to retrieve feasible continuation tokens, thereby eliminating invalid semantic IDs and significantly reducing computational overhead. To further scale to large video corpora, the Trie is optimized by combining prefix-tree search with binary search, which effectively alleviates memory overhead under a massive semantic ID space.  

UniSearch has been successfully deployed in real-world industrial search scenarios. As illustrated in Figure~\ref{pic:model} (b), the deployment architecture integrates four key components into a unified system. First, a reinforcement learning (RL) training system continuously fine-tunes UniSearch with online preference alignment and synchronizes updated parameters to the inference service. Second, an online inference system serves real-time search requests, where models are deployed using TensorRT and accelerated by key--value (KV) caching to ensure low-latency responses. Third, the Trie server provides efficient constrained generation by maintaining valid semantic ID paths, thereby guaranteeing both accuracy and scalability. Finally, a reward system evaluates candidate results through fine-grained ranking scores and user interaction signals, feeding back preference-aligned rewards to guide further training.  

Through this end-to-end design, UniSearch achieves scalable and low-latency inference while maintaining the ability to continuously improve search quality via online learning and preference optimization.

\section{Experiments}
In this section, we present the datasets, evaluation metrics, implementation details, and both offline and online experimental analyzes of UniSearch.

\subsection{Implementation Details}
\noindent \textbf{Datasets.}  
To evaluate the effectiveness of UniSearch, we construct large-scale training and testing datasets based on real search logs from the Kuaishou App. For live search, we collect approximately 80 million user search sessions spanning the period from July 9 to July 23, 2025. Data from July 9 to July 22 are used for training, while logs from July 23 form the held-out test set. All offline experiments reported in Sections~4.2 and~4.3 are conducted on this live dataset. Following the same procedure, we also construct a dataset for short-video search, which is used in Section~4.4 to validate the generalization of UniSearch beyond live scenarios.  

\noindent \textbf{Details of the UniSearch.}
The UniSearch framework consists of a Search Generator and a Video Encoder, instantiated using BART~\cite{lewis-etal-2020-bart} and BERT~\cite{devlin-etal-2019-bert} architectures, respectively. For the live search task, where the candidate pool contains around 500K live sessions, we employ lightweight 6-layer BART and BERT models with a hidden size of 768. The item encoder discretizes video embeddings into semantic IDs using a three-level codebook ($k=3$), with each level containing 512 entries. For the short-video search task, the candidate pool scales up to roughly one billion videos, which requires a larger model to handle the substantially increased complexity. In this setting, we adopt a 12-layer UniSearch configuration with a codebook of three levels and 8192 entries per level. We use the Adam optimizer with a batch size of 64. During pre-training, the initial learning rate is set to $2\times10^{-4}$ and scheduled with a cosine decay strategy. For online post-training, we use a fixed learning rate of $1\times10^{-5}$. During inference, UniSearch employs beam search with a beam size of 256 to efficiently generate high-quality semantic IDs. This setup balances computational efficiency with the ability to explore diverse candidate items, ensuring robust retrieval performance under both live-streaming and short-video search scenarios.

\noindent \textbf{Evaluation Metrics.}
To evaluate the performance of our model, we adopt Recall@300 and Mean Reciprocal Rank (MRR) as evaluation metrics. Recall@300 assesses the model’s ability to retrieve relevant results, while MRR reflects its ranking ability. For Recall@300, let $T_i$ denote the number of true positive instances among the top 300 retrieved results (\textit{i.e.}, instances that are both predicted and labeled as positive), let 
$P_i$ denote the total number of positive instances in the dataset, and let 
$N$ represent the total number of samples: 
\begin{align}
\textnormal{Recall@300} = \frac{1}{N} \sum_{i=1}^{N} \frac{T_i}{P_i}.
\end{align}
In the MRR metric, $K_i$ denotes the position of the ground-truth result for the $i$-th sample in the ranked list of candidates:
\begin{align}
\textnormal{MRR} = \frac{1}{N} \sum_{i=1}^{N} \frac{1}{K_i}.
\end{align}
We conduct experiments on both the {\it ranking subset} (RK) and  {\it click subset} (CK). The RK test set includes the top videos composed to users by the search system, used to evaluate the alignment between generated videos and system preferences. The CK test set treats clicked items as positives, demonstrating immediate user preferences.

\subsection{Main Results}
To verify the effectiveness of UniSearch, we conduct extensive experiments on the live search dataset and compare it with existing generative search frameworks. All experiments are performed under the same training and testing conditions to ensure a fair comparison.  

\begin{table*}[t]
    \caption{Performance comparison of different model on live-streaming search dataset.}
    \label{tab:results}
    \centering
    \newcolumntype{P}[1]{>{\centering\arraybackslash}p{#1}}
    \renewcommand{\arraystretch}{1.1}
    \begin{tabular}{P{1.25cm}P{2.25cm}P{2.25cm}P{2.25cm}P{1.15cm}P{1.15cm}P{1.15cm}P{1.15cm}}
    \toprule
 &     \multicolumn{3}{c}{\textbf{Model}}&\multicolumn{2}{c}{\textbf{Recall@300(\%)}$\uparrow$}& \multicolumn{2}{c}{\textbf{MRR(\%)}$\uparrow$}\\
 \cmidrule(r){2-4} \cmidrule{5-6} \cmidrule(l){7-8}
 &     \textbf{Embed. Model}&\textbf{Codebook}&\textbf{Gen. Model}&\textbf{RK}&\textbf{CK}& \textbf{RK}&\textbf{CK}\\
 \midrule
 \multirow{5}{*}[-0.75ex]{\rotatebox{90}{\shortstack{\textbf{Existing} \\ \textbf{Generative} \\\textbf{Search Arch.}}}}& & FSQ& & 55.42& 62.26& 8.36&9.93\\
 & BERT-6 Layer& VQ-VAE& BART-~6 Layer & 64.83& 63.67& 7.98&9.50\\
 &     &RQ-Kmeans&&65.56&64.12& 9.24&14.81\\
 \cmidrule{2-8}
 &     BERT-12 Layer &\multirow{2}{*}{RQ-Kmeans}&BART-12 Layer&67.79& 67.90& 12.13& 15.68\\
  &     BERT-24 Layer& &BART-24 Layer&69.05&69.27& 13.84& 16.80\\
  \midrule
  \multirow{5}{*}[-0.75ex]{\rotatebox{90}{\shortstack{\textbf{UniSearch} \\ \textbf{(Ours)}}}}&     \multicolumn{3}{c}{UniSearch-~6 Layer}&67.45&68.17&9.65&15.36\\
          &     \multicolumn{3}{c}{UniSearch-12 Layer}&68.83&70.11&12.35&  15.73\\
          &     \multicolumn{3}{c}{UniSearch-24 Layer}&69.78&72.03&14.62&  16.02\\
\cmidrule{2-8}
 & \multicolumn{3}{c}{UniSearch-~6 Layer (w/ Online SPO)}& 68.29& 68.48& 12.19&16.04\\
 & \multicolumn{3}{c}{UniSearch-12 Layer (w/ Online SPO)}& 69.53& 70.25& 13.56& 16.97 \\
 \bottomrule
 \end{tabular}
\end{table*}

\noindent \textbf{Baselines.}  
As introduced in the Introduction, existing generative search methods are not fully end-to-end. They typically involve a two-stage pipeline: (1) training an embedding model and constructing a codebook using methods such as VQ-VAE \cite{van2017neural}, FSQ \cite{mentzer2023finite}, or RQ-Kmeans; and (2) training a generative model on the tokenized items. We reproduce these approaches as baselines, and the results are reported in Table~\ref{tab:results} (the first five rows). Although such frameworks achieve reasonable performance (\textit{e.g.}, a 6-layer baseline with RQ-Kmeans reaches an MRR of 14.81 on the CK test set), their reliance on multiple models and inconsistent training objectives often leads to suboptimal results.  

\noindent \textbf{Advantages of Unified Pre-training.}  
In contrast, UniSearch unifies feature representation learning, codebook construction, and generative modeling within a single pre-training framework. This joint optimization eliminates the inconsistency across objectives and enables stronger query understanding and higher-quality generation. As shown in Table~\ref{tab:results}, UniSearch consistently outperforms baseline methods in offline evaluations. Under the same model scale, UniSearch-6 Layer significantly surpasses the 6-layer baselines in terms of MRR and achieves Recall@300 performance close to that of 12-layer baselines. This demonstrates that a unified objective enables the model to generate more relevant results while improving overall generation quality. Furthermore, UniSearch exhibits strong scalability: both UniSearch-12 Layer and UniSearch-24 Layer continue to achieve substantial improvements over baselines of the same scale, highlighting the framework’s robustness at larger capacities.  

\noindent \textbf{Effect of Search Preference Optimization (SPO).}  
We further enhance UniSearch with Search Preference Optimization (SPO), applying it to both 6-layer and 12-layer models. The results are presented in the last two rows of Table~\ref{tab:results}. With SPO, UniSearch achieves moderate gains in Recall@300 but exhibits substantial improvements in MRR. This indicates that SPO effectively guides the model toward generating results that are not only relevant but also of higher quality and better aligned with user preferences. In particular, SPO encourages the model to prioritize items that are more likely to be favored by users, such as popular or high-quality videos, thereby improving the overall user experience.

\subsection{Ablation Study}
We conduct ablation studies to evaluate the impact of several core design choices in UniSearch, focusing on the training scheme, codebook configuration, and inference strategy. 

\noindent \textbf{Effectiveness of  Coarse-to-Fine Strategy and Residual Contrastive Learning.} 
To disentangle the contributions of unified pre-training, we compare three ablated variants of UniSearch, with results summarized in Table~\ref{tab:results_abla}. The analysis is as follows.  

(1) {\it UniSearch-Plain} treats each token independently and optimizes them toward the same objective. This setup weakens token-level diversity, which ultimately yields the weakest performance on both Recall@300 and MRR.  
(2) {\it UniSearch-Plain w/ CF} augments the model with a coarse-to-fine (CF) task. By gradually learning from simpler to more complex objectives, the model gains stronger ranking capability and achieves noticeable improvements in recall and accuracy. However, because the coarse and fine tasks are inherently correlated, the resulting token representations often become similar. This may cause the \textit{path collapsing problem}, where multiple candidate items correspond to a single decoding path, thereby reducing discriminative power and harming precision.  
(3) {\it UniSearch-Plain w/ RCL} incorporates residual contrastive learning (RCL). Instead of optimizing toward the final objective in a single step, RCL progressively refines representations through residual summation. This encourages different tokens to capture complementary semantic aspects, mitigating path collapse and substantially boosting the ranking metric MRR.  

Our full UniSearch model integrates both CF and RCL. The CF strategy provides a structured learning path from easy to difficult objectives, while the residual design preserves token diversity and prevents collapse. Together, these components enable more effective end-to-end discretization, reduce clustering loss, and ultimately yield superior overall performance compared to any single variant.

\begin{table}[t]
\centering
\renewcommand{\arraystretch}{1.1}
\caption{Analysis results of Coarse-to-Fine Strategy and Residual Contrastive Learning, where CF denotes Coarse-to-Fine, RCL denotes Residual Contrastive Learning, and UniSearch-Plain refers to UniSearch without CF and RCL.}
\begin{tabular}{l*{4}{c}}
\toprule 
\multirow{2}{*}[-0.75ex]{\textbf{Method}} & \multicolumn{2}{c}{\textbf{Recall@300}(\%) $\uparrow$ } & \multicolumn{2}{c}{\textbf{MRR}(\%) $\uparrow$ }\\
\cmidrule(r){2-3} \cmidrule(l){4-5}
 & \textbf{RK} & \textbf{CK} & \textbf{RK} & \textbf{CK} \\
\midrule 
UniSearch-Plain & {63.54} & {63.15} & {6.55} & {12.16} \\
UniSearch-Plain w/ CF & {64.63} & {65.61} & {7.01} & {12.97} \\
UniSearch-Plain w/ RCL & {66.18} & {65.41} & {8.59} & {14.71} \\
\midrule 
\textbf{UniSearch} & \textbf{67.45} & \textbf{68.17} & \textbf{9.65}  & \textbf{15.36} \\
\bottomrule 
\end{tabular}
\label{tab:results_abla}
\end{table}

\begin{figure}[t]
\centering
\includegraphics[width=8cm]{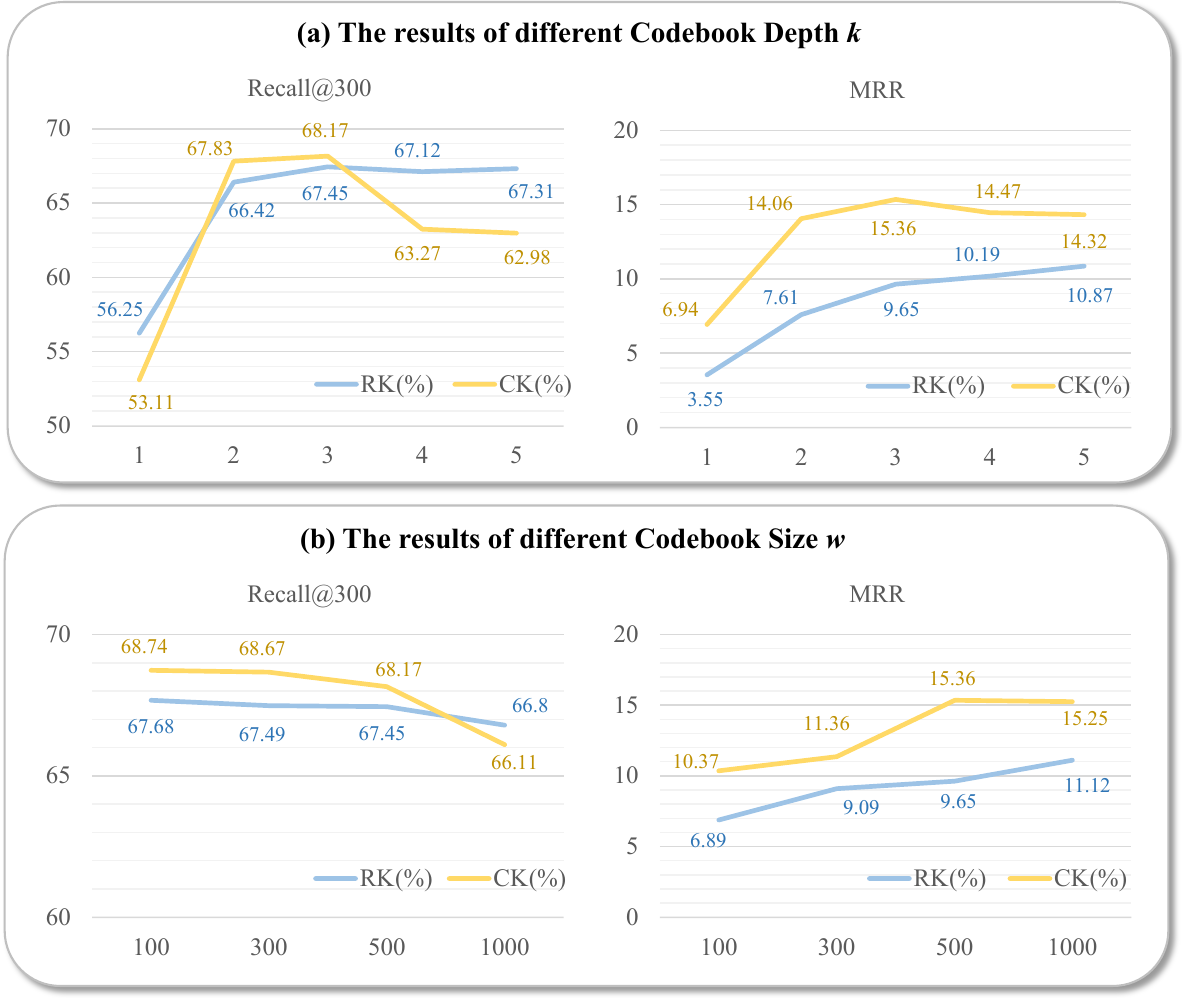}
\caption{The results of different Codebook Depth and Codebook Size on RK and CK test set. (a) The results of different Codebook Depth $k$, and (b) The results of different Codebook Size $w$.
}
\label{pic:codebook_ana}
\end{figure}

\noindent \textbf{Ablation on Codebook Depth and Size.}  
Codebook depth and size are critical hyperparameters in candidate tokenization, directly influencing retrieval quality and system efficiency. We systematically analyze their effects as follows.  

(1) To study the impact of depth, we fix the codebook size at 512. Intuitively, deeper codebooks should better capture semantic information by generating longer token sequences. However, as shown in Figure~\ref{pic:codebook_ana}(a), performance saturates once the depth $k$ exceeds 3, with both Recall@300 and MRR showing diminishing returns. A plausible explanation is that deeper codebooks increase dispersion in the candidate space, thereby weakening generalization. In addition, greater depth inevitably prolongs decoding and increases inference latency. Considering both accuracy and efficiency, we adopt a depth of $k=3$.  
(2) Next, we fix the depth at $k=3$ and vary the size of each codebook layer. The results in Figure~\ref{pic:codebook_ana}(b) reveal divergent trends: Recall@300 decreases as size grows, while MRR steadily improves. Further analysis suggests that larger codebooks encourage finer semantic partitioning, which enhances ranking precision but reduces recall by fragmenting candidate coverage. Since practical systems require a balance between recall and precision, we set the codebook size to $w=512$, which achieves strong performance across both metrics.  

\noindent \textbf{Impact of the Trie Constraint.}  
The validity of generated paths (\textit{i.e.}, whether the paths correspond to actual candidate items) is critical for system reliability. Without constraints, invalid generations may lead to empty result sets, posing significant risks to search availability. To mitigate this, we incorporate a Trie-based constraint during inference.  

\begin{figure}[t]
\centering
\includegraphics[width=8cm]{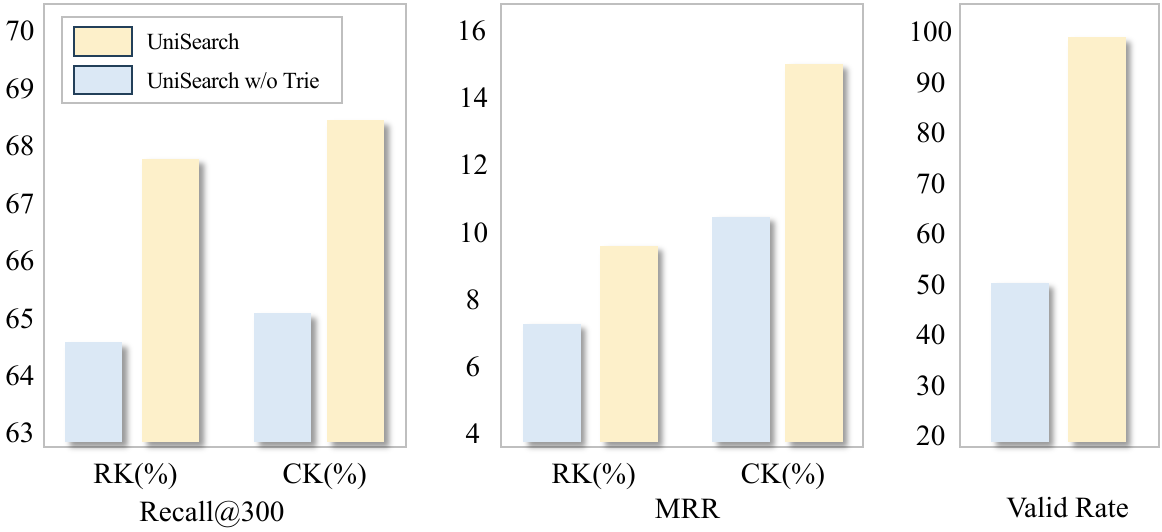}
\caption{The results of incorporating the Trie on RK and CK test set, and Valid Rates of path to candidates.}
\label{pic:exp_trie}
\end{figure}

As shown in Figure~\ref{pic:exp_trie}, the introduction of the Trie markedly improves both Recall@300 and MRR, while path validity increases from 51.3\% to 99.8\%. This ensures that nearly all generated results map to valid candidates, enhancing both reliability and computational efficiency. In live search scenarios, dynamic updates to the Trie are also essential: as live streams start or end, the set of valid paths evolves, and even for the same stream, changing content can alter its valid identifiers. Maintaining an up-to-date Trie is thus crucial to guaranteeing robustness in real-world deployments.

\begin{table}[t]
\centering
\renewcommand{\arraystretch}{1.1}
\caption{The improvements of UniSearch in online A/B test compared to the production baseline.}
\begin{tabular}{l*{4}{c}}
\toprule 
\multirow{2}{*}[-0.75ex]{\textbf{UniSearch - Live}} & \textbf{TPC} $\uparrow$ & \textbf{CTR} $\uparrow$ & \textbf{CQR} $\downarrow$ & \textbf{FCP} $\downarrow$ \\
\cmidrule(lr{0pt}){2-5}
 & +3.31\%  &  +0.202\%  &  -0.382\% & -0.107\% \\
\midrule 
\multirow{2}{*}[-0.75ex]{\textbf{UniSearch - Video}} & \textbf{VPD} $\uparrow$ & \textbf{PVD} $\uparrow$ & \textbf{CQR} $\downarrow$ & \textbf{LPC} $\uparrow$ \\
\cmidrule(lr{0pt}){2-5}
 & +0.213\%  &  +0.993\%  &  -0.602\% & +0.830\%\\

\bottomrule 
\end{tabular}
\label{tab:oneline_res}
\end{table}

\subsection{Online Testing}
\label{sec:online_test}

We further evaluate UniSearch in real-world settings by deploying it to Kuaishou’s live-streaming and short-video search systems, and conduct  online A/B tests against the legacy production system.   

\noindent \textbf{Online A/B Testing Setup.} 
We conduct a 7-day online A/B test, where both the experimental and control groups are randomly assigned 10\% of actual search traffic. For the live-streaming search scenario, we directly replace the entire Multi-stage Cascading Architecture (MCA) with UniSearch, thereby providing a clean comparison between the unified generative framework and the traditional multi-stage pipeline. In contrast, the short-video search scenario carries the highest traffic volume across Kuaishou. To ensure stability and reliability during deployment, we adopt a progressive integration strategy: UniSearch is first introduced as a new resource in the ranking stage, while several recall sources are gradually deactivated. This allows us to evaluate UniSearch’s effectiveness under strict production constraints without risking user experience.  

\noindent \textbf{Impact on Business Metrics.}  
We focus on business metrics that directly reflect user satisfaction and engagement, and summarize the results in Table~\ref{tab:oneline_res}.  
\emph{In the live-streaming search scenario}, UniSearch achieves a +3.31\% improvement in total play counts (TPC), a +0.202\% increase in page click-through rates (CTR), and reductions of -0.382\% and -0.107\% in change query rates (CQR) and position of first-click (PFC), respectively. Among these, TPC is the most critical metric, as it directly reflects search scale and platform vitality. The +3.31\% gain represents the most significant improvement in recent years. The CTR increase indicates stronger user engagement, while decreases in CQR and PFC suggest that UniSearch produces more satisfactory results with fewer reformulations and earlier clicks.  
\emph{In the short-video search scenario}, UniSearch leads to a +0.213\% improvement in video playback duration (VPD), a +0.993\% increase in page view depth (PVD), a -0.602\% reduction in CQR, and a +0.830\% increase in long play count per PV (LPC). This reflects that UniSearch not only satisfies immediate search needs but also encourages users to explore a wider variety of content, leading to more diversified interests.  

\begin{figure}[t]
\centering
\includegraphics[width=8.3cm]{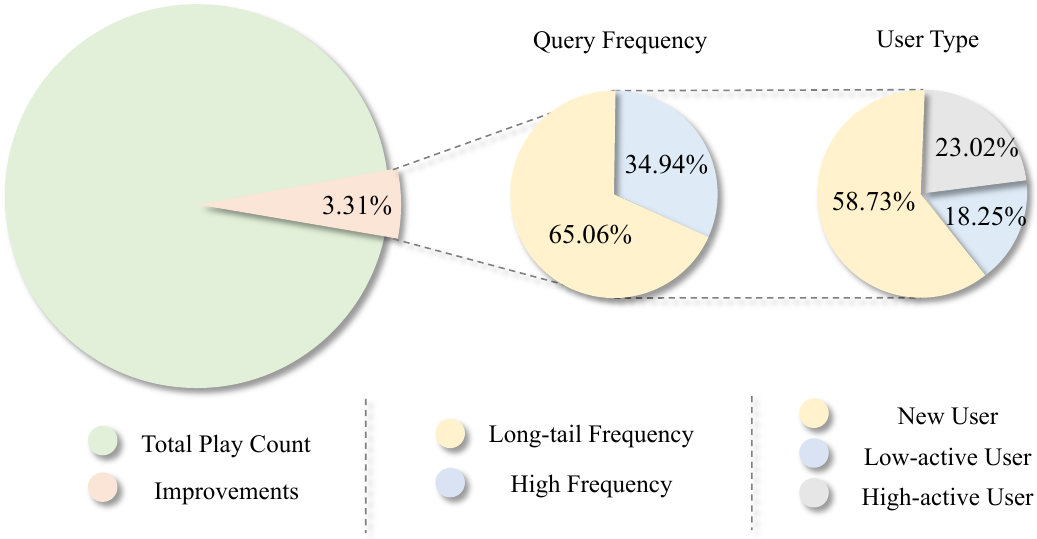}
\caption{The analysis of the improvements in Total Play Count (TPC) from query frequency and user type.
}
\label{pic:living_ana}
\end{figure}

\subsection{Further Analysis}
\label{sec:analysis}

From the online experiments reported in Section~\ref{sec:online_test}, UniSearch achieves substantial improvements in key business metrics for both live-streaming and short-video search. To better understand the underlying reasons, we perform a detailed analysis of the Live Search scenario, examining the contributions of query frequency, user activity status, and the diversity of generated results.  

\noindent \textbf{Enhanced Performance on Long-tail Queries.}  
Given that TPC is the most critical metric for Live Search, we conduct a fine-grained decomposition of its gain. As shown in Figure~\ref{pic:living_ana}, 65.06\% of the TPC improvement is driven by long-tail queries—nearly double the contribution from head queries. This underscores the importance of effectively modeling long-tail traffic in industrial search. Long-tail queries are typically characterized by their sparse occurrence and semantic complexity, making them challenging for traditional systems like MCA to handle. UniSearch, by contrast, benefits from unified pre-training and improved semantic representations, enabling it to capture subtle user intents and generalize more effectively across diverse and low-frequency queries.  

\noindent \textbf{Stronger Appeal to New Users.}  
We further segment the TPC gains by user activity status. As illustrated in Figure~\ref{pic:living_ana}, 58.73\% of the increase originates from new users, accounting for more than half of the total gain. This suggests that UniSearch delivers results that are particularly attractive to users who are less familiar with the platform. A plausible explanation is that existing users have already adapted to the behavioral patterns of the legacy MCA system, while new users are more sensitive to improved semantic understanding and richer result sets. This improvement in new-user engagement is especially valuable, as it directly contributes to expanding the active search user base and sustaining long-term growth.  

\begin{figure}[t]
\centering
\includegraphics[width=8cm]{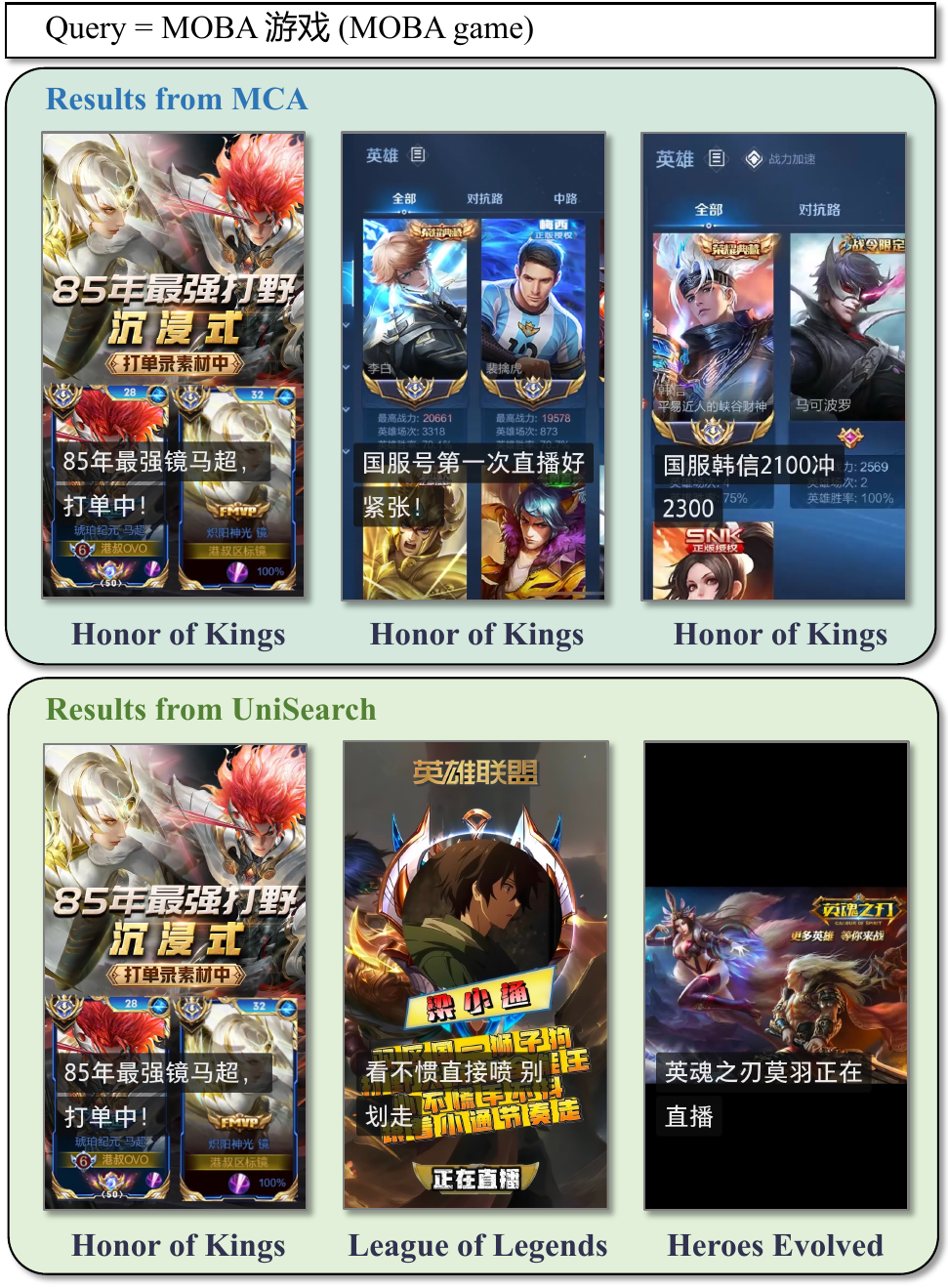}
\caption{Search results comparison between the Multi-stage Cascading Architecture (MCA) and our UniSearch.
}
\label{pic:case}
\end{figure}

\noindent \textbf{Richer and More Diverse Results.}  
Case studies further validate that UniSearch generates results that are both semantically relevant and diverse. For example, Figure~\ref{pic:case} presents the query ``MOBA game''. Under MCA, the top three results across users are consistently ``Honor of Kings'', showing strong bias toward a single popular item. In contrast, UniSearch surfaces additional relevant games such as ``League of Legends'' and ``Heroes Evolved''. This broader coverage ensures that users with different preferences all find satisfactory content. Without such diversity, users whose interests are not reflected (\textit{e.g.}, those preferring ``League of Legends'') may abandon the platform and reduce future engagement. By balancing relevance with diversity, UniSearch not only improves user satisfaction but also mitigates the risk of content homogenization.

\section{Conclusion}
We proposed \textbf{UniSearch}, a unified generative search framework that replaces the traditional multi-stage cascaded architecture with an end-to-end solution. By jointly optimizing a Search Generator and an Video Encoder, UniSearch integrates tokenization and generation within a single training paradigm, leading to improved representation learning and generation quality. In addition, the proposed \emph{Search Preference Optimization} leverages online reward system and user feedback to align generation with user preferences.  
Extensive offline experiments and online A/B testing on large-scale industrial datasets demonstrate that UniSearch outperforms strong baselines in both retrieval quality and efficiency. Its deployment in Kuaishou live-streaming search achieved the most significant metric improvement in recent years, highlighting its practical value.  

\noindent \textbf{Limitations and Future Work.}  
Currently, UniSearch generates candidates in a \emph{point-wise} manner using beam search, which may limit result diversity. Future work will focus on enhancing \emph{list-wise generation} to improve both diversity and ranking accuracy, as well as exploring finer-grained reward methods to further strengthen alignment with user preferences.

\bibliographystyle{ACM-Reference-Format}
\bibliography{sample-base}



 \end{document}